\documentstyle[aps,epsf]{revtex}

\title{QUASIELASTIC ELECTRON SCATTERING FROM NUCLEI:\\
RANDOM-PHASE VS. RING APPROXIMATIONS}

\author{Eduardo Bauer$^{1,}$ \cite{AAAuth} 
and Antonio M. Lallena$^2$}

\address{$^1$Departamento de F\'{\i}sica, Facultad de Ciencias
Exactas, Universidad Nacional de La Plata, La Plata, 1900,
Argentina.\\
$^2$Departamento de F\'{\i}sica Moderna, Universidad de
Granada, E-18071 Granada, Spain.}

\begin{document}
\maketitle

\vspace{.2cm}

\begin{abstract}
We investigate the extent to which the nuclear transverse
response to electron scattering in the quasielastic region,
evaluated in the random-phase approximation can be described by
ring approximation calculations. Different effective
interactions based on a standard model of the type
$g'+V_\pi+V_\rho$ are employed. For each momentum transfer, we
have obtained the value of $g_0^\prime$ permitting the ring
response to match the position of the peak and/or the non-energy 
weighted sum rule provided by the random-phase
approach has been obtained. It is found that, in general, it is
not possible to reproduce both magnitudes simultaneously for a
given $g_0^\prime$ value.

\vspace{.1cm}


\end{abstract}

\section{INTRODUCTION}

In the past years, much attention has been paid to the study of the
electron scattering (nuclear) responses in the quasi-free regime. 
The good description of the cross sections
provided by means of a simple Fermi gas model in the former work of
Moniz {\it et al.}\/ \cite{mo71} was suddenly broken when the
longitudinal/transverse experimental separation was performed
\cite{expe}. After this, many different physical mechanisms, such as,
e.g., short- and long-range correlations, meson-exchange currents,
final state interactions, etc., have been argued to be responsible of 
the observed discrepancies. However, a definite answer to the problem
is still not available.

Calculations of the nuclear responses in this energy region can be
grouped in two general approaches. A first one considers the nucleus
as a finite system \cite{ca84}-\cite{sl95}. The other one uses
nuclear matter together with an additional approximation (say, a
variable Fermi momentum or the local density approximation) to obtain
the results for finite nuclei \cite{al84}-\cite{ce97}. 

Nuclear matter formalism takes advantage of the translational
invariance inherent to the infinite systems, something which
simplifies considerably the technology to be used (at least, {\it a
priori}). However, most of the calculations done in this approach have
been performed in the so-called ring approximation (RA)
\cite{al84,ba95}-\cite{ce97}. This framework is usually (and
incorrectly) called random-phase approximation (RPA), though the
exchange terms are not considered. Curiously, full true RPA nuclear
responses have been evaluated only for finite nuclei \cite{bu91},
despite the complexity of the calculations for these systems in
comparison with those for nuclear matter. A first attempt to
carry out RPA calculations for infinite systems was done in Ref.
\cite{de96}, where the longitudinal response was evaluated by
means of the continued fraction method with exchange terms considered
up to first order only. More recently, 
two different procedures to calculate the nuclear matter responses 
in a RPA framework have been developed for a general finite range 
effective interaction \cite{ba96,de98}.

It is commonly assumed \cite{os82} that the RA can simulate the effect
of RPA exchange terms by an adequate choice of the Landau parameters
included in the interaction. In particular, for the transverse
responses, in which we are interested in this work, the $g_0^\prime$
parameter will be the important one. However, Shigehara {\it et al.}\/
\cite{sh89} have shown that this is true in the $\sigma \tau$ response
for finite nuclei when a particular $G$-matrix, which has a weak
momentum dependence in the exchange channel is used as effective
interaction. The validity of this hypothesis for the standard
$g_0^\prime + V_\pi + V_\rho$ model has not been clarified. 

This is precisely the aim of the present investigation: the study of
the possibility for the RA to describe RPA calculations with such an 
interaction. In Sec.~II we compare the RPA responses with the RA ones
in order to obtain the values of $g_0^\prime$ providing the best
agreement between both. In Sec.~III we go deeper in the question by
analyzing the results obtained for two effective interactions obtained
by slightly modifying the one used in the previous section.
Finally, we present our conclusions in Sec.~IV.

\section{RPA vs. RA}

We start by performing a ``model'' RPA calculation for the
quasielastic nuclear response in $^{40}$Ca. We are interested in the
transverse channel and we have used an effective interaction of the
form

\begin{figure}[ht]
\vspace*{-3.4cm}
\begin{center}                                                                
\leavevmode
\epsfysize = 400pt
\hspace*{.45cm}
\makebox[0cm]{\epsfbox{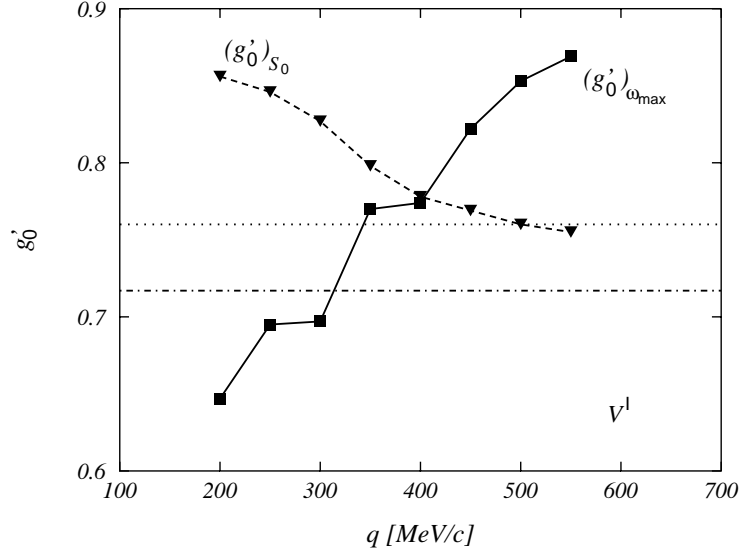}}
\end{center}
\vspace*{-3.5cm}
\caption{Dependence with the momentum transfer $q$ of the
values of the parameter $g_0^\prime$ to be used in RA
calculations in order to reproduce the peak positions (squares)
and the the non-energy weighted sum rule (triangles) corresponding 
to the RPA responses. The dotted
line gives the value $g_0^\prime=0.76$ used in the RPA
calculation. The dashed-dotted line shows the value
$g_0^\prime=0.717$ which permits to reproduce the low-energy
properties we consider by means of a RA calculation (see Ref.
\protect\cite{ba98}).}
\end{figure}

\vspace{.5cm}

\begin{equation}
\displaystyle
V^{\rm I} \, = \, V_{\rm LM} \, + \, V_\pi \, + \, V_\rho 
\, , 
\end{equation}
which includes a zero-range force of Landau-Migdal type, which takes
care of the short-range piece of the NN interaction: 
\begin{equation}
V_{\rm LM} \, = \,C_0 \,[ g_0\, 
{\mbox{\boldmath $\sigma$}}(1)\cdot {\mbox{\boldmath $\sigma$}}(2) 
+\, g_0^\prime\,
{\mbox{\boldmath $\sigma$}}(1)\cdot {\mbox{\boldmath $\sigma$}}(2)
{\mbox{\boldmath $\tau$}}(1)  \cdot {\mbox{\boldmath $\tau$}}(2)]
\, ,
\end{equation}
and a finite-range component generated by the ($\pi$+ $\rho$)-meson
exchange potentials. The particular values of the two parameters of
the zero-range piece are $g_0=0.47$ and $g_0^\prime=0.76$ (with
$C_0=386$~MeV~fm$^3$). These values permit to reproduce, within the
RPA framework, the energies and B-values of the two $1^+$ states in 
$^{208}$Pb at 5.85 and 7.30~MeV. These are the (low-energy)
observables we consider to fix the different interactions we use 
throughout this work (see Ref.~\cite{ba98} for details).

For the calculation of the RPA nuclear responses in the quasi-free
region, we have used the prescription of the scheme developed
in Ref. \cite{ba96} and in which 
the exchange terms are explicitly taken into account
for any interaction. For a pure contact interaction exchange
terms can be included up to infinite order, while for a finite
range one they must be numerically evaluated for each order.  

We want to investigate the conditions under which the RA responses 
provide a reasonable description of the RPA ones. The difference
between both approaches is in the presence (or not) of the exchange
terms, which are linked to the finite range piece of the
interaction. Then we maintain fixed this part of $V^{\rm I}$ in the RA
calculations and vary the value of $g_0^\prime$ until the required
agreement is obtained. This agreement will be ``measured'' by comparing
the values obtained in both approaches for two magnitudes derived from
the corresponding responses: the position of the peak
$\omega_{\rm max}$ and the non-energy weighted sum rule
\begin{equation}
\label{SR}
\displaystyle
S_0(q) \, = \, \int_0^\infty \, {\rm d}\omega \, S_{\rm T}(q,\omega)
\, ,
\end{equation}
where $S_{\rm T}$ is the structure function corresponding to point 
nucleons, that is without including the corresponding nucleon form
factor. If the full transverse response $R_{\rm T}$ is used in
Eq. (\ref{SR}) instead of $S_{\rm T}$, the results quoted below
remain unchanged. 
We call $\left( g_0^\prime \right)_{\omega_{\rm max}}$ 
and $\left( g_0^\prime \right)_{S_0}$, respectively, the values of the 
parameter $g_0^\prime$ which make the values of $\omega_{\rm max}$ and
$S_0$ obtained within the RA equal the RPA ones.

In Fig.~1 we show the results obtained in this procedure for momentum
transfers ranging from 200 to 550~MeV/$c$. Therein, the black squares 
represent the values $\left( g_0^\prime \right)_{\omega_{\rm max}}$,
whereas the solid triangles correspond to 
$\left( g_0^\prime \right)_{S_0}$. The dotted line gives the 
$g_0^\prime$ value used in the RPA calculation. We have not changed the
value of $g_0$ because, as shown in Ref. \cite{ba98}, its role in the
RA is negligible. 

\newpage

\begin{figure}[ht]
\vspace*{-1.5cm}
\begin{center}                                                                
\leavevmode
\epsfysize = 400pt
\hspace*{.45cm}
\makebox[0cm]{\epsfbox{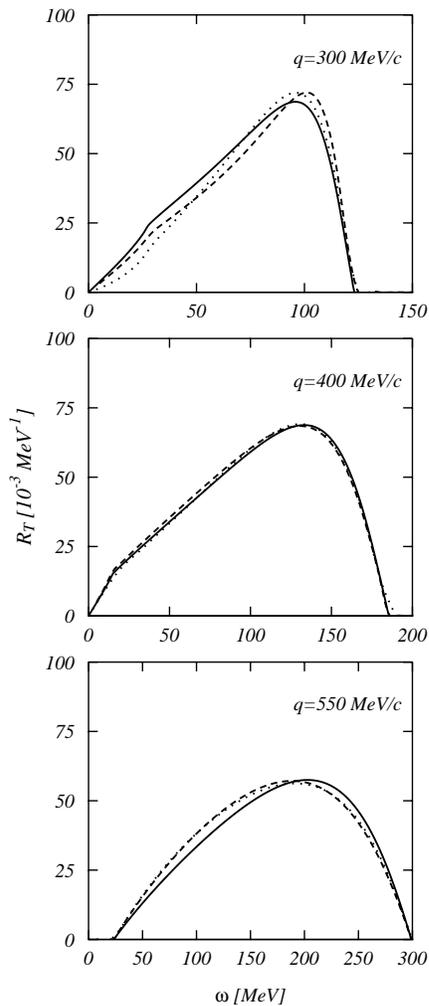}}
\end{center}
\vspace{-0.cm}
\caption{Transverse responses for $^{40}$Ca, calculated
for three momentum transfers. Dotted curves are the RPA results.
Solid curves represent the RA responses obtained with the values
$\left( g_0^\prime \right)_{\omega_{\rm max}}$.  Dashed curves
give the same but with the values $\left( g_0^\prime
\right)_{S_0}$. The particular values used in the RA
calculations are given in Table~I.}
\end{figure}

\vspace{.6cm}

\begin{table}
\caption{Values of the $g_0^\prime$ parameters used in the RA
calculations shown in Fig~2.}
\begin{tabular}{ccccc}
$q$~[MeV/$c$] & ~~~~& $\left( g_0^\prime \right)_{\omega_{\rm max}}$
& ~~~~& $\left( g_0^\prime \right)_{S_0}$\\ \hline
300. && 0.697 && 0.827 \\
400. && 0.774 && 0.778 \\
550. && 0.869 && 0.755 \\
\end{tabular}
\end{table}

\vspace{.6cm}

\begin{table}
\caption{Adjusted values of the parameters $g_0$ and 
$g_0^\prime$ for RPA calculations using the two interactions quoted in
the text. With these values and with $C_0=386$~MeV~fm$^3$)
the energies and B-values of the two $1^+$ 
states in $^{208}$Pb at 5.85 and 7.30~MeV, respectively, are
reproduced.}
\begin{tabular}{ccccc}
Interaction & ~~~~&$g_0$ & ~~~~&$g_0^\prime$ \\ \hline
$V^{\rm II}$ && -0.055 && 0.64 \\
$V^{\rm III}$ && -0.075 && 0.60 \\
\end{tabular}
\end{table}

\newpage

These results deserve several comments:
\begin{enumerate}
\item It is clear that the reproduction of the RPA values of
$\omega_{\rm max}$ and $S_0$ by means of the RA calculations occurs
for values of $g_0^\prime$ which are, in general, quite different from
that used for the RPA calculation (dotted line). This is in agreement
with the findings of Ref. \cite{ba98}.

\item The $g_0^\prime$ values permitting the agreement between both 
types of calculations for the magnitudes taken into account are 
clearly incompatible. Only the region around $q=400$~MeV/$c$ seems to
be ``magic'' in this respect. This is also seen in Fig.~2 where we
show the transverse responses for $q=300$ (upper panel), $q=400$
(medium panel) and $q=550$~MeV/$c$ (lower panel) obtained in the RPA 
(dotted curves), in the RA with 
$\left( g_0^\prime \right)_{\omega_{\rm max}}$ (solid curves) and with
$\left( g_0^\prime \right)_{S_0}$ (dashed curves). Is is apparent how 
the three curves overlap in the case of $q=400$~MeV/$c$, while they
differ in the other two cases. This result generalizes those found by
Shigehara {\it et al.}\/ for a $G$-matrix interaction \cite{sh89}.

\item The value of the $g_0^\prime$ parameter needed to obtain the
agreement between RA and RPA shows a considerably
dependence on the momentum transfer $q$, the range of variation being
appreciably large. Besides, the values providing the agreement
between both type of calculations are (except for a couple of values
around 300~MeV/$c$) quite different from the value of
$g_0^\prime=0.717$ (dashed-dotted line in Fig.~1) found \cite{ba98} 
to provide, in the RA framework, the description of the low-energy
properties quoted above. This points out even more the difficulties
for the RA to reproduce the RPA results in the quasielastic region.
\end{enumerate}

\section{ADDITIONAL RESULTS}

The results quoted in the previous section show the inability of the
RA calculations to describe the responses obtained in the RPA
framework. To go deeper in the investigation of the reasons
of this situation, we focus our
attention in the exchange terms and in those mechanisms providing the
more important contributions to them. In particular we will analyze,
first, the role of the pion exchange potential and, second, the
importance of the tensor piece of the interaction.

As it is known, the contribution of $V_\pi$ to the RA responses is
exactly zero in nuclear matter, while the same does not occurs for the
RPA because of the presence of the exchange terms. In order to see
what is the influence of this piece of the potential,
we have performed a new set of calculations, similar to the previous
ones, but considering the effective interaction:
\begin{equation}
\displaystyle
V^{\rm II} \, = \, V_{\rm LM} \, + \, V_\rho 
\, .
\end{equation}

Following the same strategy as for $V^{\rm I}$, we have fixed the
values of the zero-range Landau-Migdal parameters $g_0$ and
$g_0^\prime$ as indicated above. The results obtained are given in the
first row of Table~II.

With the interaction fixed in this way we have obtained the
corresponding RPA responses and have determined, again, the values of
$g_0^\prime$ making the RA results to agree with the RPA ones.
The results obtained are shown in the upper panel of Fig.~3. 

The most important question to be noted is the fact that the absence 
of the pion exchange potential in the RPA calculations strongly
modifies the situation. In fact, it can be seen that, in the $q$ region
between 300 and 500~MeV/$c$, a value for $g_0^\prime\sim 0.5$ would 
provide RA calculations describing reasonably well and simultanoeusly,
both $\omega_{\rm max}$ and $S_0$ as given by the
RPA. This is shown in Fig.~4 where we compare, for the interaction 
$V^{\rm II}$ we are discussing, the RA responses obtained for 
$g_0^\prime=0.505$ (solid curves) with the RPA ones (dotted curves). 
This value of $g_0^\prime$ is the one which makes RA and RPA
calculations to coincide at $q=300$~MeV/$c$ and it is worth to point
out the big difference with repect to the value $g_0^\prime=0.64$
used for the RPA calculations (see Table~II). 

In order to know more about the behavior of the important pieces of
the interaction, we have repeated the analysis done for $V^{\rm I}$
and $V^{\rm II}$ for the effective force:
\begin{equation}
\displaystyle
V^{\rm III} \, = \, V_{\rm LM} \, + \, (V_\rho)_{\sigma \sigma \tau \tau} 
\, ,
\end{equation}
which has been obtained by eliminating 
the pion exchange and the tensor piece of the $\rho$-exchange from
$V^{\rm I}$. The adjustment of the zero-range parameter at
low-energy (as in the two previous calculations) gives the values
quoted in the second row of Table~II. The results for the values of
$\left( g_0^\prime \right)_{\omega_{\rm max}}$ and 
$\left( g_0^\prime \right)_{S_0}$ are shown in the lower panel of
Fig.~3. The situation now is roughly the same as for $V^{\rm II}$, but
for a smaller value of $g_0^\prime$. These results show the importance
of the role of the pion exchange potential in this type of calculations. 

It should be also noted that, as it occurs in the case of $V^{\rm II}$, 
the $g_0^\prime$ value used for the RPA calculations differs from those
needed for the RA ones. This claims again the necessity of
changing the values of the zero-range parameters fixed in the RPA
framework when performing calculations in a different framework, something
which is not usually done in the literature.
\newpage

\begin{figure}[ht]
\vspace*{-2.cm}
\begin{center}                                                                
\leavevmode
\epsfysize = 400pt
\hspace*{.45cm}
\makebox[0cm]{\epsfbox{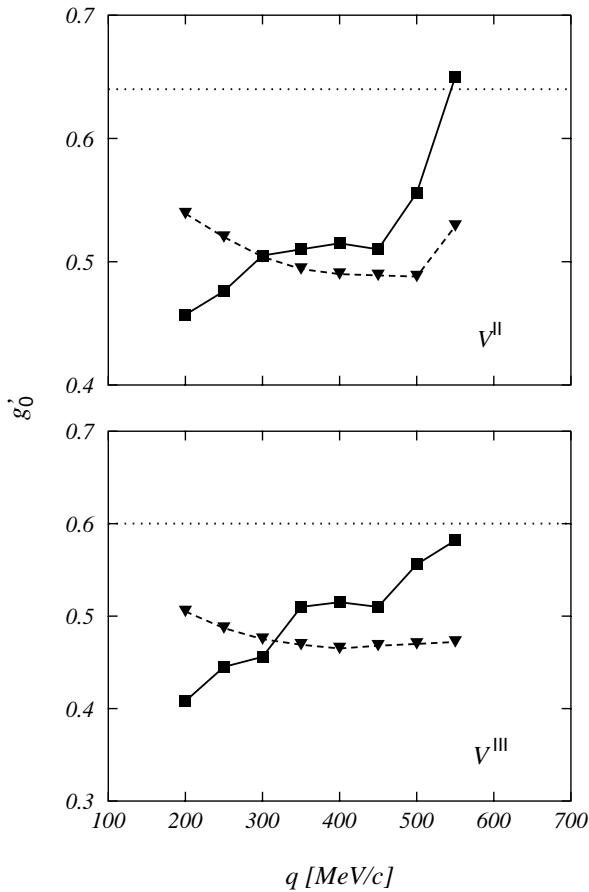}}
\end{center}
\vspace{-.5cm}
\caption{Same as in Fig.~1 but for the other two
interactions considered in this work.}
\end{figure}

\section{CONCLUSIONS}

In this work we have addressed the role played by the RPA exchange
terms in the (e,e') nuclear response in the quasielastic region.
In particular, we have investigated if the RA calculations performed
with an effective interaction with a fix $g_0^\prime$ (independent
of the momentum transfer) can simulate the results obtained in the
RPA. The main findings are the following:
\begin{enumerate}
\item It is not possible to find a single $g_0^\prime$ value
permitting the RA to reproduce the RPA responses. The required 
$g_0^\prime$ shows a strong $q$ dependence. Besides, this dependence
is different when different properties of the responses are considered
to match the results obtained with the two approaches. As a
consequence, it can be concluded that the RA cannot reproduce the RPA
responses in a consistent way. 

\item It is important to stress that pion exchange does not
contributes to the RA calculations in the transverse channel. It was
found that if $V_\pi$ is arbitrarily turn off in the effective 
interaction used for RPA calculations, then a reasonable agreement
between both approaches is obtained for $300$~MeV/$c \leq q \leq
500$~MeV/$c$. This shows the important role played by this part of the
interaction in the type of caculations we have discussed here.

\end{enumerate}

\acknowledgements

We are grateful to G. Co' for valuable
discussions. One of us (E.B.) acknowledges the warm hospitality
extended to him while visiting the Universidad de Granada.
This work has been supported in part by the Agencia Nacional de
Promoci\'on Cient\'{\i}fica y Tecnol\'ogica (Argentina) under contract
PMT-PICT-0079, by Fundaci\'on Antorchas (Argentina), by the DGES
(Spain) under contract PB95-1204 and by the Junta de Andaluc\'{\i}a 
(Spain). 

\newpage 

\begin{figure}[ht]
\vspace*{-1.cm}
\begin{center}                                                                
\leavevmode
\epsfysize = 400pt
\hspace*{.45cm}
\makebox[0cm]{\epsfbox{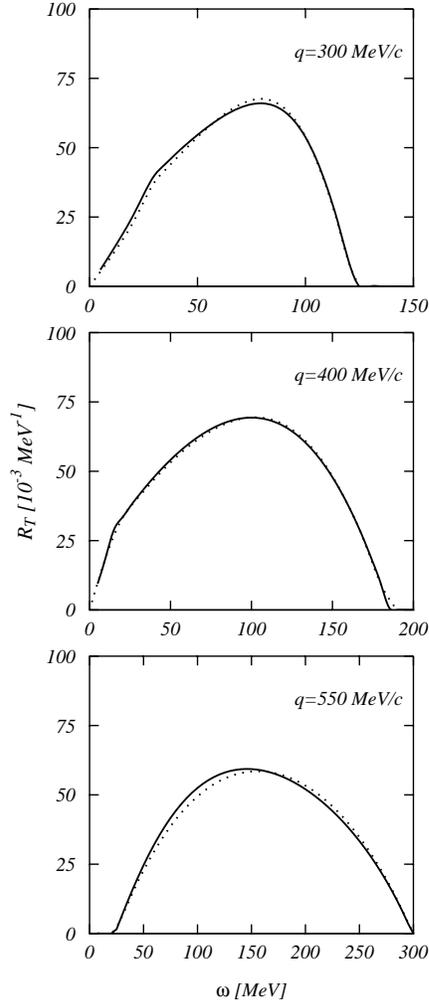}}
\end{center}
\vspace{-0.cm}
\caption{Transverse responses for $^{40}$Ca, calculated
for three momentum transfers. Dotted curves are the RPA results.
Solid curves represent the RA responses obtained with the value
$g_0^\prime=0.505$.}
\end{figure}

\end{document}